# Tunability of Planar Photonic Crystal Nanocavity: A Comparative study

Ashfaqul Anwar Siraji
*Department of Electrical and Computer Engineering
Wayne State University
Detroit, Michigan, 48201
Email: fj8972@wayne.edu

*Abstract*—A comparative study of the tunability of planar photonic crystal nanocavities is done in this work. Three different types of cavities, e.g., defect cavity, double heterostructure cavity and bandedge cavity are studied using finite domain time domain method. Electrically tunable materials like Graphene and ferroelectric $BaTiO_3$ is used as the active materials. We found that high quality factor cavities have lower tunability and greater loss in the material leads to greater tunability.

*Keywords—Tunability, $BaTiO_3$, Graphene, Planar Photonic Crystal Cavity*

## I. Introduction

Planar photonic crystal (PhC) based optical nanocavities offer extremely high spectral purity, which makes them useful for sensing applications, lasers etc [1]–[5]. Aside from the high Q/V value, PhC nanocavities bring a few other advantages. The resonant properties of a PhC nanocavity can be tuned by changing the refractive index of the base material. Faraon et al. has reported a method to locally tune the resonant frequency of a 2-D PhC resonator by depositing a layer of chalcogenide glass on a GaAs PhC [6]. Dorjgotov et al. has proposed a highly tunable one dimensional PhC which consists of alternating layers of liquid crystal and dielectric material [7]. Dayal et al. has proposed a PhC with tunable bandgap by using ferroelectric materials [8]. Ferroelectrics are suitable material for realizing electro-optic tunability due to their high electro-optic coefficient. Barium titanate is a particularly attractive optical ferroelectric due to its relatively low absorption in optical range and nonlinear optical properties. Lin et al. have fabricated and characterized a PhC waveguide on $BaTiO_3$ thin films [9] and simulated a voltage tunable PhC wide band filter which can be tuned in a 7 THz range using the electro-optic property of $BaTiO_3$ [10]. We have designed several tunable PhC nanocavity using $BaTiO_3$ as the base material [11]-[13].

Graphene is another viable option for realizing electro-optic tunable PhC resonators. Gan et al. demonstrated that transferring a single layer of Graphene on a nanophotonic res- onator enables selective, orders of magnitude enhancement of optical coupling with Graphene [14]. Majumdar et al. showed that efficient tunability of nanophotonic resonators can be achieved by coupling them to Graphene [15]. Majumder et. al. also demonstrated a tunable Silicon photonic resonator where tunability is implemented by simply putting an electrically gated layer of Graphene on top of the resonator [16]. Gan et. al. also demonstrated high contrast electro-optic modulation of photonic crystal resonators by putting an electrically gated layer of Graphene on top of the resonator [17]. So far, a comparative study of tunable PhC nanocavity using electro-optic tunability of Graphene and Barium Titanate has not been reported in literature. The Ferroelectric $BaTiO_3$, is an electooptically active material. Its refractive index can be changed by an applied electric field. The change in the refractive index of $BaTiO_3$ can be given as [18]:

$$n(E) = n_0 + \Delta n \qquad (1)$$

In this paper, We have computed the resonant properties of the designed cavities with $\Delta n = 0.02$ to investigate the tunability of BaTiO3 based devices. We have also calculated the resonant properties of Graphene based devices as a function of gate voltage. We have used finite difference time domain (FDTF) method with perfectly matched layer (PML) boundary condition to determine the tunable resonance of different planar PhC nanocavities based on the two materials respectively. A comparison of the results is shown.

## II. Method

To find the resonant frequency of the planar PhC nanocavities, they are excited with a large range of frequencies. The excitation is allowed to propagate freely. Most of the frequencies do not excite the cavity and their energy leaves the cavity quickly. The resonant frequencies excite the cavity and their energy decay very slowly. Hence, the resonant frequencies come as sharp peak in the frequency spectrum. A 2-D FDTD method is used to carry out the calculations. The initial field is a broadband impulse applied at the core. The Yee's mesh is used for discretization of the simulation domain in both space and time, thus reducing the Maxwell's curl equation into simple difference equations. The space grid is setup uniformly so that there are exactly 24 grid points per period, which provides sufficiently converged results. A time step $\Delta t = 5.20833 \times 10^{-17}$s is used, which is many times smaller than the Courrant limit, but produces sufficient frequency resolution at the expense of simulation time. The material properties of $BaTiO_3$ is taken from [19] and that of Graphene is calculated using the formalism in [12].

We have studies a defect cavity [11], a double heterostructure cavity [12] and a defectless bandedge cavity in this work [13]. They are shown in Fig. 1.

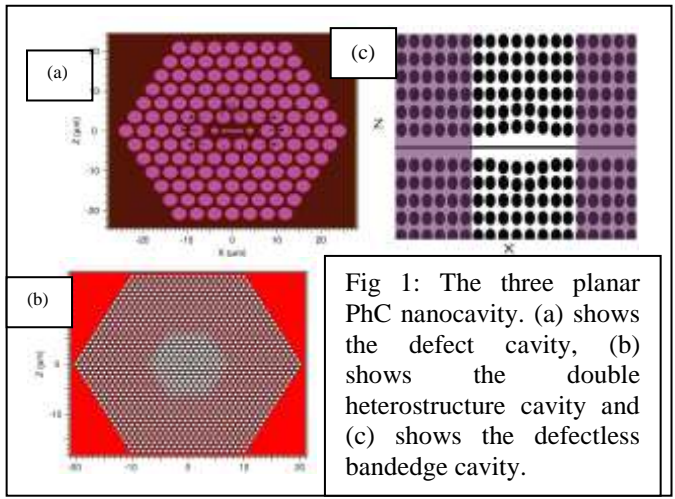

Fig 1: The three planar PhC nanocavity. (a) shows the defect cavity, (b) shows the double heterostructure cavity and (c) shows the defectless bandedge cavity.

### III. RESULT

#### A. Defect Cavity

We have mentioned previously that the base material in our calculation is BaTiO3 which is electooptically active. Its refractive index can be changed by an applied electric field. To investigate the effect of applied electric field on the properties of the designed cavity, we performed simulations with step increase in the refractive index, $\Delta n = 0.02$. The effect of the change in electric field and hence the change in refractive index on the frequency spectrum of the proposed cavity is shown in Fig. 2. It is evident that the resonant frequency of the cavity changes by 89 nm with $\Delta n = 0.02$. This shift is significant compared to the range of 24 nm in a photonic crystal resonator proposed in [20]. The sensitivity of the cavity to change in refractive index is measured by the wavelength sensitivity, $S_\lambda = \Delta \lambda_r / \Delta n$. The impact of space modulation on the wavelength sensitivity of the L3 cavity and the proposed cavity are plotted in Fig. 3. It can be seen that the sensitivity of both the cavities depend on modulation depth. The dependence is linear in case of the L3 cavity, but sublinear in case of the proposed cavity.

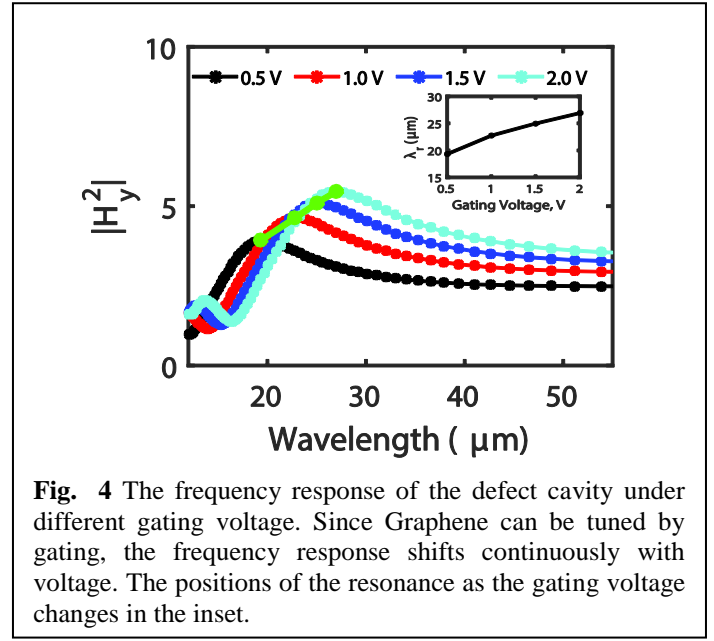

**Fig. 4** The frequency response of the defect cavity under different gating voltage. Since Graphene can be tuned by gating, the frequency response shifts continuously with voltage. The positions of the resonance as the gating voltage changes in the inset.

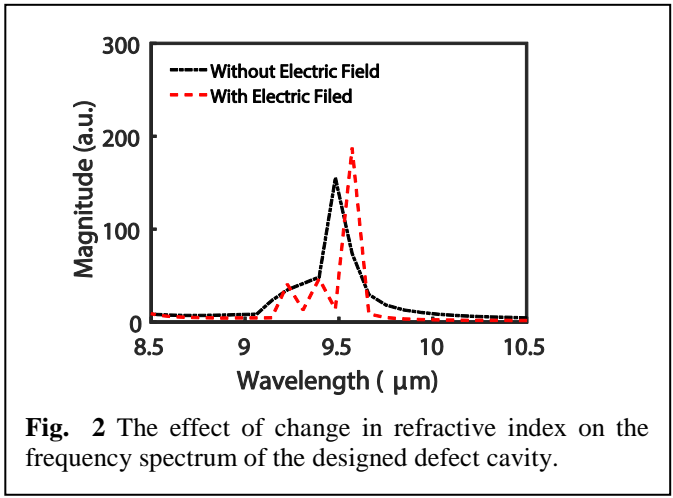

**Fig. 2** The effect of change in refractive index on the frequency spectrum of the designed defect cavity.

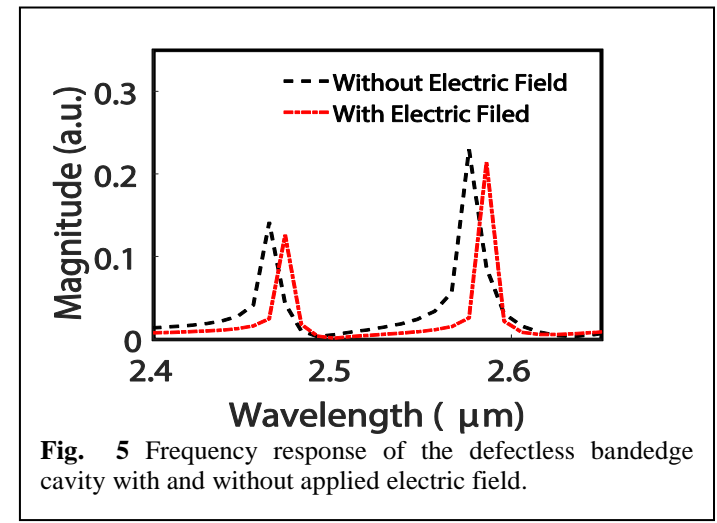

**Fig. 5** Frequency response of the defectless bandedge cavity with and without applied electric field.

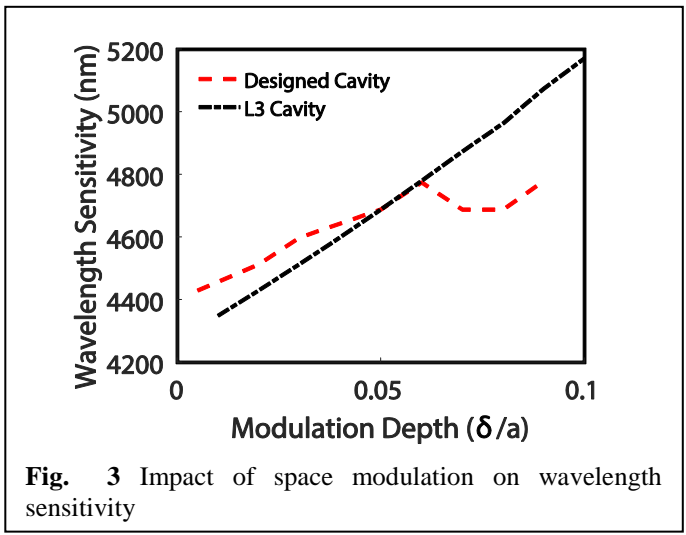

**Fig. 3** Impact of space modulation on wavelength sensitivity

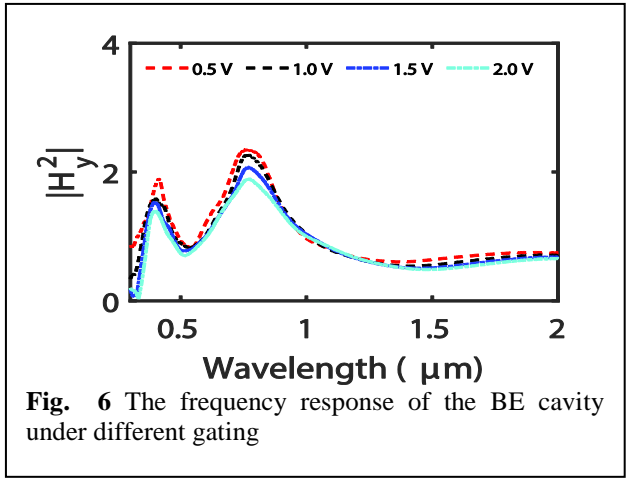

**Fig. 6** The frequency response of the BE cavity under different gating

When Graphene is used as the base materials for the defect cavity, the quality factor of the cavity decreases. However, the optical properties of Graphene are readily tunable by gating. Hence, the defect cavity becomes much more tunable when Graphene is used as the base material. In the Fig. 4, the impulse response of the cavity is shown with different gating voltage. It can be seen that the impulse responses shift continuously under the influence of progressively higher gating voltage. As shown in the inset, the resonant wavelength shifts from 19.32 µm to 26.95 µm as the gating voltage changes from 0.5 V to 2V. Evidently, the range of tuning is much higher at 7.63 µm when Graphene is used, as opposed to the tuning range of 89 nm when $BaTiO_3$ is used.

### B. Defectless Bandedge Cavity

The studied structure has two resonant TE modes, which are separated by more than 100 nm. The difference between the resonant frequencies indicate that they are not degenerate modes separated by mesh coarseness. The resonant modes are increasingly further from the dielectric band-edge of the mirror PhC.

We calculated the frequency response of the cavity with and without applied electric field, as shown in Fig. 5. It can be seen that the resonant peaks are slightly moved to higher wavelengths when an electric field in applied. It can be seen that the resonant modes are moved to higher wavelength by 10 nm and 11 nm respectively. Although this change demonstrates dynamic tunability of the device, it is much weaker than changes in resonant wavelengths mentioned in case of the defect cavity.

When Graphene is used instead of $BaTiO_3$, the flexible tunability of Graphene enhances the tunability of the BE cavity, but only slightly. From Fig. 6, it can be seen that both resonant peaks continuously shift under increasing gating voltage. Here, we demonstrated a tuning range of 15 nm for the first mode, which is higher than that obtained when the base material is $BaTiO3$. However, this range is still lower than the range obtained in case of a defect based cavity. This illustrates that the low tuning range of BE cavity is independent of material selection. Rather it is due to the high spectral purity of the resonances in BE cavity. The strong resonances occur for the slow modes at the edge of the dielectric band, which always remain within a small band of wavelengths, despite small changes to the optical properties.

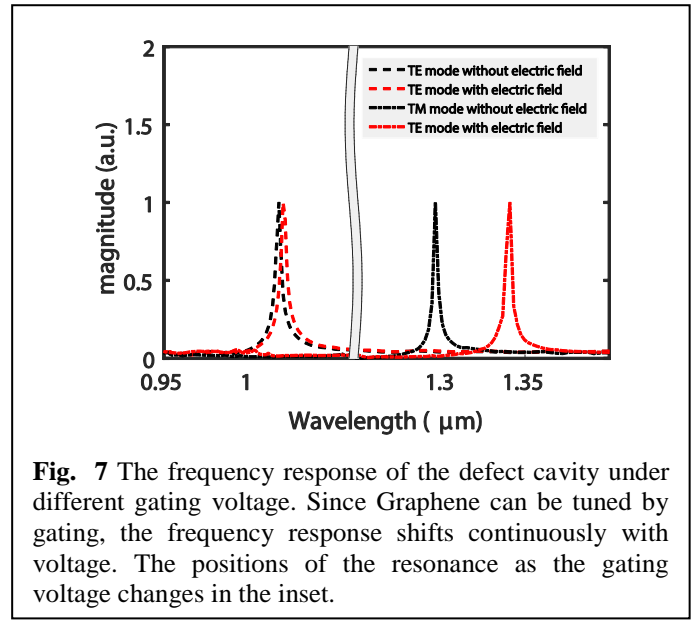

**Fig. 7** The frequency response of the defect cavity under different gating voltage. Since Graphene can be tuned by gating, the frequency response shifts continuously with voltage. The positions of the resonance as the gating voltage changes in the inset.

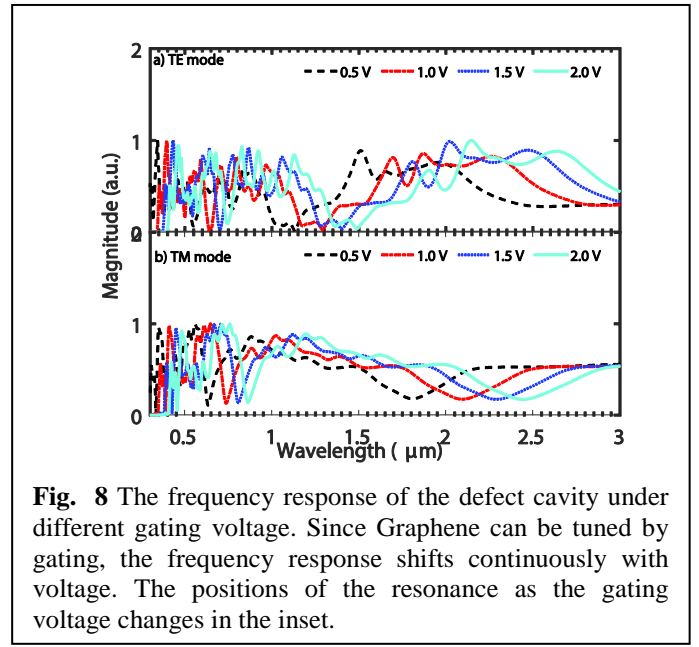

**Fig. 8** The frequency response of the defect cavity under different gating voltage. Since Graphene can be tuned by gating, the frequency response shifts continuously with voltage. The positions of the resonance as the gating voltage changes in the inset.

### C. Double-Heterostructure cavity

To approximate the change in refractive index under applied bias, we assume a step increase in the refractive index as done in [18]. In Fig. 7, the effect of applied electric field on the frequency response of TE and TM modes of the cavity is shown. For both cases, the resonant frequency shifts as electric field in applied. This shift is 45 nm for the TM mode and 2 nm for the TE mode. The effect is significant for the TM mode compared to that of 24 nm shown in [15]. Thus, the cavity shows better tunability of its TM modes compared to its TE modes.

In Fig. 8, the frequency response of the DH cavity for the TE and TM modes, respectively, are shown when the base material is Graphene. It can be seen that, for the TE mode, there exists a distinct peak which shifts continuously with the applied gating voltage. The peak for the TE mode shifts from

1.509 µm to 2.145 µm, a range of 636 nm, which is many times higher than the range when BaTiO3 was used as a base material. The Frequency response of the TM mode also shifts by 135 nm as the gating voltage is shifted from 0.5 V to 2.0 V. However, for the TM mode, the peaks are weak and not very clear. Since the TM modes aren't well confined, the interaction with the TM mode and the Graphene base material is weak, leading to weak tunability. However, the TE modes are distinct and strongly confined, leading to stronger interaction between the modes and Graphene base material. Hence the tunability of TE modes are higher.

The results are summarized in Table 1.

**Table 1: Summary of the results**

| Nanocavity | Mode | Q | λ (µm) | $d\lambda_{Bto}$(nm) | $d\lambda_{Gp}$(nm) |
|---|---|---|---|---|---|
| Defect Cavity | TE | 4062 | 9.48 | 89 | 7630 |
| Bandedge Cavity | TE | 4132 | 2.465 | 10 | 15 |
|  | TE | 4098 | 2.576 | 11 | 15 |
| DH Cavity | TE | 2190 | 1.019 | 2 | 636 |
|  | TM | 2800 | 1.296 | 45 | 135 |

## IV. CONCLUSION

We have compared the tunability of three different planar photonic crystal nanocavities with BaTiO$_3$ and Graphene as the active material. We have found that the defect cavity shows greatest tunability whereas double heterostructure cavities shows smallest tunability. This is because the confinement of DH cavity is due to mode gap where as that of defect cavity is due to bandgap. The modes of BE cavity shows significant tunability compared to the DH cavity. Furthermore, it was seen that using lossy material like Graphene led to a greater tuning range compared to low loss BaTiO$_3$. The tunability analysis is useful for choosing appropriate structures for different application. Forexample a fixed wavelength light source is best implemented using a DH cavity whereas a multicolor laser is best implemented using a defect cavity.

REFERENCES

[1] Y. Zhang, M. Khan, Y. Huang, J. Ryou, P. Deotare, R. Dupuis, and M. LonCar, "Photonic crystal nanobeam lasers," Appl. Phys. Lett., vol. 97, no. 5, p. 051104, Aug. 2010.
[2] D.-U. Kim, S. Kim, J. Lee, S.-R. Jeon, and H. Jeon, "Free-standing GaN-based photonic crystal band-edge laser," IEEE Photon. Technol. Lett., vol. 23, no. 20, pp. 1454–1456, Oct. 2011.
[3] F. Gonzalez and J. Alda, "Optical nanoantennas coupled to photonic crystal cavities and waveguides for near-field sensing," IEEE J. Sel. Topics Quantum Electron., vol. 16, no. 2, pp. 446–449, Apr. 2010.
[4] X. Gan, N. Pervez, I. Kymissis, F. Hatami, and D. Englund, "A high-resolution spectrometer based on a compact planar two dimensional photonic crystal cavity array," Appl. Phys. Lett., vol. 100, no. 23, p. 231104, June 2012.
[5] Y. Li, K. Cui, X. Feng, Y. Huang, D. Wang, Z. Huang, and W. Zhang, "Photonic crystal nanobeam cavity with stagger holes for ultrafast directly modulated nano-light-emitting diodes," IEEE Photon. J., vol. 5, no. 1, p. 4700306, Feb. 2013.
[6] A. Faraon, D. Englund, D. Bulla, B. Luther-Davies, B. J. Eggleton, N.Stoltz, P. Petroff and J. VuCkovic, "Local tuning of photonic crystal cavities using chalcogenide glasses," Appl. Phys. Lett., vol. 92, no. 4, 2008.
[7] E.-A. Dorjgotov, A. Bhowmik and P. J. Bos, "High tunability mixed order photonic crystal," Appl. Phys. Lett., vol. 96, no. 16, 2010.
[8] K. Dayal and K. Bhattacharya, "Active tuning of photonic device characteristics during operation by ferroelectric domain switching," J. Appl. Phys., vol. 102, no. 6, 2007.
[9] P. T. Lin, F. Yi, S.-T. Ho, and B. Wessels, "Two-dimensional ferroelectric photonic crystal waveguides: Simulation, fabrication, and optical char- acterization," J. Lightw. Technol., vol. 27, no. 19, pp. 4330–4337, Oct. 2009.
[10] P. T. Lin, Z. Liu, and B. W. Wessels, "Ferroelectric thin film photonic crystal waveguide and its electro-optic properties," J. Opt. A Pure Appl. Op., vol. 11, no. 7, p. 075005, May 2009.
[11] A.A. Siraji, M.S. Alam, S. Haque, "Impact of Space Modulation on Confinement of Light in a Novel Photonic Crystal Cavity on Ferroelectric Barium Titanate," *Journal of Lightwave Technology*, vol.31, no.5, pp.802,808, March, 2013.
[12] A.A. Siraji, M.A. Alam, "A Tunable Photonic Double Heterostructure Cavity on Ferroelectric Barium Titanate," IEEE Photonics Technology Letters, vol.25, no.17, pp.1676,1679, Sept.1, 2013.
[13] A.A. Siraji, M.A. Alam, "Design of a tunable high Q photonic band edge cavity on ferroelectric Barium Titanate," *Advances in Electrical Engineering (ICAEE), 2013 International Conference on*, vol., no., pp.48,53, 19-21 Dec. 2013.
[14] X. Gan, R.-J. Shiue, Y. Gao, S. Assefa, J. Hone and D. Englund, "Controlled Light-Matter Interaction in Graphene Electrooptic Devices Using Nanophotonic Cavities and Waveguides," IEEE J. Sel. Topics Quantum Electron., vol. 20, no. 1, pp. 95-105, Jan.-Feb. 2014.
[15] A. Majumdar, J. Kim, J. Vuckovic and Feng Wang, "Graphene for Tunable Nanophotonic Resonators," IEEE J. Sel. Topics Quantum Electron., vol. 20, no. 1, pp. 68-71, Jan.-Feb. 2014.
[16] A. Majumdar, J. Kim, J. Vuckovic, and F. Wang, "Electrical control of silicon photonic crystal cavity by graphene," Nano Letters, vol. 13, no. 2, pp. 515–518, 2013.
[17] X. Gan, R.-J. Shiue, Y. Gao, K. F. Mak, X. Yao, L. Li, A. Szep, D. Walker, J. Hone, T. F. Heinz, and D. Englund, "High-contrast electrooptic modulation of a photonic crystal nanocavity by electrical gating of graphene," Nano Letters, vol. 13, no. 2, pp. 691–696, 2013.
[18] M. J. Dicken, L. A. Sweatlock, D. Pacifici, H. J. Lezec, K. Bhattacharya, and H. A. Atwater, "Electrooptic modulation in thin film Barium Titanate plasmonic interferometers," Nano Lett., vol. 8, no. 11, pp. 4048–4052, 2008.
[19] A.A. Siraji, M.A. Alam, "Improved calculation of the electronic and optical Properties of tetragonal barium titanate." *Journal of electronic materials* 43.5 (2014): 1443-1449.
[20] X. Chew, G. Zhou, F. S. Chau, and J. Deng, "Nanomechanically tunable photonic crystal resonators utilizing triple-beam coupled nanocavities," IEEE Photon. Technol. Lett., vol. 23, no. 18, pp. 1310–1312, Sept. 2011.